%% file: mplex-genrep.tex
\documentclass[sigconf]{aamas}  

\usepackage{balance} 
\usepackage{changepage}
\usepackage{dcolumn}
\usepackage{tikz}
\usepackage{pgfplots}
\usepackage{subfig}
\usepackage{amsmath}
\usepackage{amssymb}
\usepackage{pgf}
\usetikzlibrary{arrows,automata}
\usepackage[latin1]{inputenc}
\usepackage{verbatim}
\usepackage{physics}
\usetikzlibrary{positioning, decorations.pathmorphing}

\definecolor{mynavy}{HTML}{000080}
\definecolor{darkred}{HTML}{8B0000}
\definecolor{mygreen}{HTML}{006400}
\definecolor{mygold}{HTML}{B8860B}

\newcolumntype{d}[1]{D..{#1}}
\pgfplotsset{compat=1.14}

\setcopyright{ifaamas}  
\acmDOI{}  
\acmISBN{}  
\acmConference[AAMAS'18]{Proc.\@ of the 17th International Conference on Autonomous Agents and Multiagent Systems (AAMAS 2018)}{July 10--15, 2018}{Stockholm, Sweden}{M.~Dastani, G.~Sukthankar, E.~Andr\'{e}, S.~Koenig (eds.)}  
\acmYear{2018}  
\copyrightyear{2018}  
\acmPrice{}  

\begin{document}

\title{Multiplex Network Structure Enhances the Role of Generalized Reciprocity in Promoting Cooperation}  



\author{Viktor Stojkoski}
\affiliation{%
  \institution{Macedonian Academy of Sciences and Arts}
  \city{Skopje} 
  \state{Republic of Macedonia} 
  \postcode{P.O. Box 428}
}
\email{vstojkoski@manu.edu.mk}
\author{Zoran Utkovski}
\affiliation{%
  \institution{Fraunhofer Heinrich Hertz Institute}
  \city{Berlin} 
  \state{Germany} 
  \postcode{Einsteinufer 37, 10587}
}
\email{zoran.utkovski@hhi.fraunhofer.de}
\author{Elisabeth Andr\'{e}}
\affiliation{%
  \institution{Augsburg University}
  \city{Augsburg} 
  \state{Germany} 
  \postcode{Universitätsstr. 6a 86159}
}
\email{andre@informatik.uni-augsburg.de}
\author{Ljupco Kocarev}
\affiliation{%
  \institution{Macedonian Academy of Sciences and Arts}
  \city{Skopje} 
  \state{Republic of Macedonia} 
  \postcode{P.O. Box 428}
}
\email{lkocarev@manu.edu.mk}
\begin{abstract}  
In multi-agent systems, cooperative behavior is largely determined by the network structure which dictates the interactions among neighboring agents. These interactions often exhibit multidimensional features, either as relationships of different types or temporal dynamics, both of which may be modeled as a "multiplex" network. Against this background, here we advance the research on cooperation models inspired by \textit{generalized reciprocity}, a simple pay-it-forward behavioral mechanism, by considering a multidimensional networked society. Our results reveal that a multiplex network structure can act as an enhancer of the role of generalized reciprocity in promoting cooperation by acting as a latent support, even when the parameters in some of the separate network dimensions suggest otherwise (i.e. favor defection). As a result, generalized reciprocity forces the cooperative contributions of the individual agents to concentrate in the dimension which is most favorable for the existence of cooperation.
\end{abstract}


\maketitle

\input{mplex-genrep_body}


\bibliographystyle{ACM-Reference-Format}  
\balance
\bibliography{mplex-bib}  

\end{document}

%% file: mplex-genrep_body.tex
\section{Introduction}

Recent biological studies suggest that cooperative behavior can emerge and be sustained if it is based on \textit{generalized reciprocity}, a rule rooted in the principle of ``help anyone if helped by someone''~\cite{Taborsky-2016}. In~\cite{Rutte-2007, Bartlett-2006, Isen-1987} it was shown that cooperation under this mechanism may be a consequence of the changes in the physiological state of the agents caused by their positive experience from previous interactions. 

The first steps towards developing a framework for studying the role of a state-based generalized reciprocity update rule in networked societies were made in \cite{Utkovski-2017,stojkoski2018cooperation}. In these works, the incentives for cooperation are determined by a sole variable called \textit{internal cooperative state} which reflects the agents' current welfare. A distinctive characteristic of the model is that, in steady state, the simple decision rule promotes cooperation while, at the same time, prevents the agents from being exploited by the environment.

While this and similar models shed valuable insights on the role that network topology plays in promoting cooperation, most of them have so far addressed only interactions on networks that are of one ``dimension'', ignoring possible multidimensional phenomena, i.e. multiplex network structures. This is a slight drawback since real-life networks often exhibit heterogeneous properties within the edge structure that are of fundamental value to the phenomena present in the system~\cite{Kivela-2014}. For instance, in social network analyses the patterning and interweaving of different types of relationships are needed to describe and characterize social structures \cite{Boorman-1976,White-1976}. In telecommunication networks, the physical edges are often ``sliced'' into multiple parts in order to support the requirement of different devices \cite{Sherwood-2010,Nikaein-2015}. Even genetic and protein relations between organisms constructed in multiple ways are crucial for the analysis of their global interaction properties~\cite{Stark-2006, DeDomenico-2015}.

To this end, here we extend the model introduced in~\cite{Utkovski-2017} to account for a multiplex network structure, with the aim to characterize the network cooperation dynamics under the premise of a state-based behavioral mechanism rooted in generalized reciprocity. In our extension the dimensions act as platforms which facilitate transactions between active members. The activity of the agents is modeled by constraining their presence to one dimension per round, and by making them able to answer only to requests from that same dimension. While this assumption is consistent with the random walk models on multiplex networks~\cite{de2014navigability}, it is additionally rational to be implemented in systems where the round duration is very short and/or when agents have limited interaction capacity. The resulting mechanism, which retains the property of preventing exploitation, exhibits additional numerical features that act as promoters of cooperation in a multiplex network structure. 

\section{Model and results}

\begin{figure*}[t!]
\includegraphics[width=14cm]{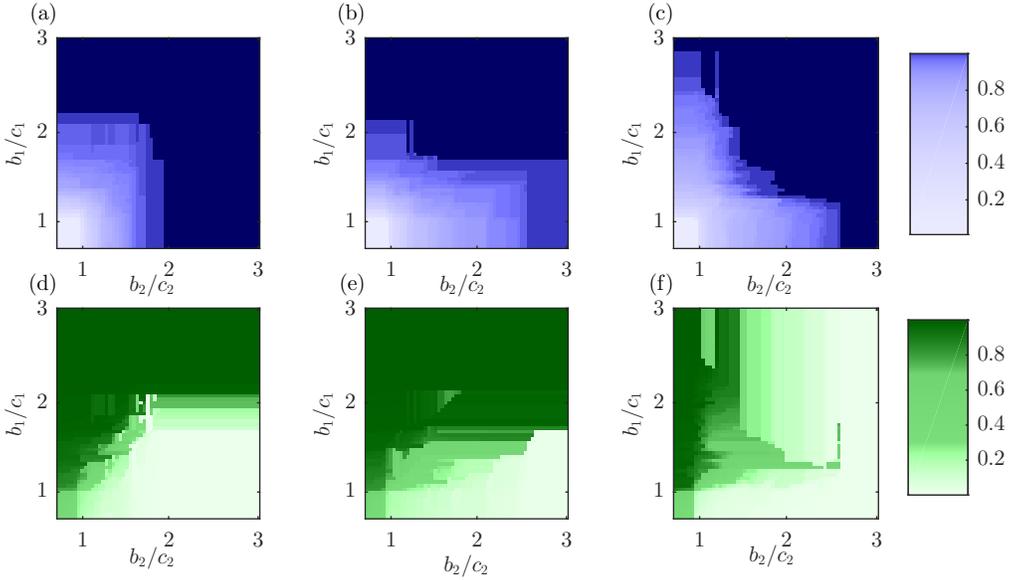}
\caption{Heat map for the fraction of unconditional cooperators as a function of the benefit and cost ratio $b_l/c_l$ for a sample of two dimensional random graphs in which there is an update rule for dimension presence. \textbf{(a)} ER-ER random graph. \textbf{(b)} ER-BA random graph. \textbf{(c)} BA-BA random graph. \textbf{(d)-(f)} Heat map for the steady state values for the average probability that an agent is present in the first dimension for the same graphs. All networks have 100 agents and average degree 8.}
\label{fig:dimension-update}
\end{figure*}

We consider a population of $N$ agents whose relations are modeled as a connected multiplex network, defined as the triplet $\mathcal{G} \left( \mathcal{N}, \mathcal{E}, \mathcal{L} \right)$, where $\mathcal{N}$ (the set of nodes) corresponds to the set of agents, $\mathcal{E} \subseteq \mathcal{N} \times \mathcal{N}$ is the set of edges that describes the relationships between pairs of agents, and $\mathcal{L}$ is the set of $L$ properties that can be attributed to the edges and which define the dimensions of the network. 

The interactions between the agents are modeled as follows: in each round $t$, each agent $i$ randomly chooses a dimension $l$ where it will be present in that round with probability $B_{il}(t)$; then it sends a cooperation request to a randomly (on uniform) chosen agent $j$ from its neighborhood in the $l$-th dimension. Upon selection, if agent $j$ is present in the the $l$-th dimension, it receives the request and cooperates with probability $\mathrm{p}_j(t)$ representing the agent's internal cooperative state at round $t$; When cooperating, agent $j$ pays a cost $c_l > 0$ for agent $i$ to receive a benefit $b_l > 0$.

 We base the behavioral update rule on the accumulated payoff of the agent $i$ by round $t$, $\mathrm{Y}_{i}(t)=\mathrm{Y}_{i}(t-1)+\mathrm{y}_{i}(t)$, with $\mathrm{Y}_{i}(0)$ being the initial condition and $\mathrm{y}_{i}(0)=0$ is the payoff in round $0$. Formally, the cooperative state of $i$ at round $t+1$ is defined as 
\begin{align}
\mathrm{p}_i(t+1)=\mathrm{f} \left[\mathrm{Y}_{i}(t)\right],
\label{eq:update}
\end{align} 
where we assume that the function $\mathrm{f}:\mathbb{R}\rightarrow [0,\:\:1]$ is increasing. A plausible choice which reflects real-world behavior is the logistic function $\mathrm{f}(\omega)=\left[1+e^{-k(\omega-\omega_0)}\right]^{-1}$,
where the parameters $k$ and $\omega_0$ define the steepness and the midpoint of the function. 

In a similar fashion, we define the dimension update rule as
\begin{align}
B_{il}(t+1) = \frac{\exp\left( \mathrm{Y}_{il}(t) \right) }{ \sum_{m} \exp\left( \mathrm{Y}_{im}(t) \right)},
\label{eq:dimension_update}
\end{align}
where $\mathrm{Y}_{il}(t)$, is the accumulated payoff in dimension $l$ by round t. 

In the numerical study of the model properties, we compare realizations of three different multiplex networks each composed of two dimensions that are generated independently. In particular, the dimensions of the first multiplex network are constructed by generating two Erdos-Renyi (ER) random graphs, the dimensions of the second by generating one Erdos-Renyi random graph and one Barabasi-Albert (BA) scale-free graph, and the third by generating two Barabasi-Albert graphs. The results are shown in Fig.~\ref{fig:dimension-update}. Panels (a)-(c) show the heat map for the fraction of individuals with steady state incentive for cooperation $\mathrm{p}^*_i$ equal to one as a function of the benefit to cost ratios. For all network types we observe that cooperation may exist even if the benefit to cost ratio in a dimension suggest otherwise. This is a result of the fact that the negative payoffs from the original dimension are compensated with positive payoffs from the other dimension. If at least one agent receives higher steady state payoff from the supporting dimension than the loss in the original, then cooperation will persist. This aggregate behavior can be explained by looking at panels (d)-(f) of the figure, where we display the heat map for the average steady state probability for being present in the first dimension as a function of the same parameters. Obviously, the dimension in which the agents are always present in steady state is not always the same, i.e. it is dispersed among the agents depending on the network topology and parameters. This is a key feature of the model since it implies that the dimension update rule forces the agents to accommodate their presence towards the dimension where they are either least burdened to cooperate or in which most of their cooperative neighbors are present. As such, when coupled with the generalized reciprocity state update rule (\ref{eq:update}), the dimension update rule eases the promotion of cooperation in the system, in the sense of existence of agents with steady states probability for cooperation $\mathrm{p}^*_i$ greater than zero.